\documentclass[10pt,pra,aps,twocolumn]{revtex4-1}

\usepackage{amsmath}
\usepackage{amsfonts}
\usepackage{amssymb}
\usepackage{graphicx}
\usepackage{amsthm} 	
\usepackage{mathrsfs}
\usepackage{wrapfig}
\usepackage{txfonts} 
\usepackage{setspace}
\usepackage{color}
\usepackage{psboxit}
\usepackage{bm}
\usepackage{bbm}
\usepackage{subfigure}
\usepackage{float}

\newcommand{\norm}[1]{\lVert#1\rVert}
\newcommand{\ket}[1]{\left|{#1}\right>}
\newcommand{\bra}[1]{\left<{#1}\right|}

\newtheorem{definition}{Definition}
\newtheorem{theorem}{Theorem}
\newtheorem{lemma}{Lemma}

\begin{document}

\title{Degree of separability of bipartite quantum states}

\author{\surname{Thiang} Guo Chuan}
\affiliation{Centre for Quantum Technologies, National University of Singapore, 3 Science Drive 2, 117543, Singapore}

\date{\today}

\begin{abstract}
	We investigate the problem of finding the optimal convex decomposition of a bipartite quantum state into a separable part and a positive remainder, in which the weight of the separable part is maximal. This weight is naturally identified with the degree of separability of the state.  In a recent work, the problem was solved for two-qubit states using semidefinite programming. In this paper, we describe a procedure to obtain the optimal decomposition of a bipartite state of any finite dimension via a sequence of semidefinite relaxations. The sequence of decompositions thus obtained is shown to converge to the optimal one. This provides, for the first time, a systematic method to determine the so-called optimal Lewenstein--Sanpera decomposition of any bipartite state. Numerical results are provided to illustrate this procedure, and the special case of rank-2 states is also discussed. 
\end{abstract}


\maketitle

\section{Introduction}
In recent years, a large amount of effort has been put into the study of quantum entanglement, driven in part by the realization of its enormous potential as a resource in quantum information processing \cite{horodecki09}. The \emph{separability problem} has received particular attention --- this asks for the determination of whether a given state of a composite system is separable or not. This basic question remains an open problem, and indeed, it has already been shown to be $NP$-hard \cite{gurvits03}. Consequently, complete operational criteria for separability are known only in special cases or low dimensions. This poses major problems for the characterization of entanglement. For one, it makes the quantification of entanglement an extremely difficult task. Many quantities of interest, such as the best separable approximation measure \cite{lewenstein98}, robustness of entanglement \cite{vidal99} and geometric measure of entanglement \cite{wei03}, involve some kind of optimization over the set of separable states, which cannot be handled easily. However, as the separable states form a convex set, the study of entanglement often leads to convex optimization problems \cite{boyd04}. One can then benefit from the tremendous advances made recently in the field of convex optimization, as well as the increasing availability of powerful computing equipment. In particular, semidefinite programming (SDP) \cite{vandenberghe96} has found its way, very naturally, into a variety of problems in quantum information theory, and has been used in the context of distillable entanglement \cite{rains01}, completely positive maps \cite{audenaert02}, entanglement witnesses \cite{brandao04}, unambiguous state discrimination \cite{eldar04}, and linear optics quantum gates \cite{eisert05}. 

In this paper, we are concerned with the Lewenstein--Sanpera decomposition \cite{lewenstein98} (LSD), which is essentially a convex decomposition of a mixed bipartite state into a separable part and a positive remainder. The optimal LSD has maximal weight on the separable part, and it is this particular decomposition that we are interested in. The matter of finding the optimal LSD is hardly trivial, and even in the two-qubit case, the optimal LSDs were only known for some special states \cite{wellens01, englert00, englert02}. In a recent paper \cite{thiang09}, it was shown that the optimal LSD problem for two-qubit states can be rephrased as a semidefinite program, effectively solving the problem for this simplest possible composite system. This is possible due to the Peres--Horodecki criterion \cite{peres96,horodecki96}, which turns a troublesome separability constraint into a positivity and therefore tractable constraint. Coupled with an analysis of the dual SDP problem, optimality conditions characterizing the optimal LSD of two-qubit states were derived. For qubit-qutrits, the same analysis can be carried out, because the Peres--Horodecki criterion for separability still holds. Generalizing this approach, we will show that the optimal LSD of any bipartite state of arbitrary finite dimension can be accessed via a sequence of semidefinite relaxations of the optimization problem. Each step in such a sequence involves solving a semidefinite program, for which numerous reliable and efficient solvers are available \cite{grant10, TTT09, sturm99, lofberg04, fujisawa05}. This procedure utilizes the separability criterion introduced by Doherty \emph{et al.} \cite{doherty02, doherty04}, and the complementary one provided by Navascu\'es \emph{et al.} \cite{navascues09, navascues09-2}, which involve searching for symmetric extensions of the state in question.

The structure of this paper will be as follows: in Section \ref{sec:review}, we introduce the notion of Lewenstein--Sanpera decompositions, and discuss the main features of semidefinite programming. In Section \ref{sec:LSDviasymm}, we present our main result --- a systematic procedure for obtaining the optimal LSD of an arbitrary bipartite state, using the tools of semidefinite programming and the symmetric extensions criterion for separability. Results from our numerical implementation will be used to illustrate the convergence properties of our scheme. Finally, in Section \ref{sec:lowrank}, we will explore the optimal LSDs of qubit-qubit and qubit-qutrit states, and then explain how one can obtain the optimal LSD of arbitrary rank-2 states analytically.

\section{Review of Lewenstein--Sanpera decompositions and semidefinite programming}\label{sec:review}

\subsection{Lewenstein--Sanpera decompositions}
Given an arbitrary bipartite quantum state $\rho\in\mathcal{B}(\mathcal{H}_A\otimes\mathcal{H}_B)$, one can look for convex decompositions of $\rho$ into a separable part and a positive remainder. Such decompositions are called Lewenstein--Sanpera decompositions \cite{lewenstein98}. Since the set of separable states is compact, an optimal decomposition in which the separable part has maximal weight certainly exists. Furthermore, it was shown by Karnas and Lewenstein that this optimal decomposition is unique \cite{karnas01} for systems of any finite dimension. We denote the optimal LSD of $\rho$ by
\begin{equation}
	\rho=\mathcal{S}\varrho_\text{sep}+(1-\mathcal{S})\varsigma_\text{ent}\equiv\tilde{\varrho}_\text{sep}+\tilde{\varsigma}_\text{ent}.\label{optlsd}
\end{equation}
In the previous equation and in what follows, calligraphic font is used to indicate quantities that are optimal, while a tilde above an operator indicates that it is not normalized to unit trace. It is natural to identify the maximal weight $\mathcal{S}$ as the \emph{degree of separability} of the state $\rho$. Finding the optimal LSD for an arbitrary state is in fact a convex constrained optimization problem, in which one maximizes $\text{tr}\{\tilde{\rho}_\text{sep}\}$, a linear objective function, over the convex cone of separable linear operators, subject to the constraint that the difference $\rho-\tilde{\rho}_\text{sep}=\tilde{\sigma}_\text{ent}$ remains positive semidefinite.

\subsection{Semidefinite programming}
The primal semidefinite program \cite{vandenberghe96} has the following form:
\begin{equation}
	\begin{array}{ll}
	\text{minimize} & \vec{c}^{\,\text{T}}\vec{x}\\
	\text{subject to} & F(\vec{x})\geq 0,\label{primalprogram}
	\end{array}
\end{equation}
where	$F(\vec{x})=F_0+\sum_{i=1}^mx_iF_i$ and $\vec{x}\in\mathbb{R}^m$. The given inputs for the primal problem are (i) the vector $\vec{c}\in\mathbb{R}^m$ characterizing the objective function, and (ii) the $m+1$ hermitian $n\times n$ matrices $F_0,F_1,\ldots,F_m$ defining the linear matrix inequality. The dual problem associated with \eqref{primalprogram} is
\begin{equation}
	\begin{array}{ll}
		\text{maximize} &-\text{tr}\{F_0Z\}\\
		\text{subject to} &\text{tr}\{F_iZ\}=c_i,\, i=1,\ldots,m,\\
		{} & Z\geq 0.
	\end{array}\label{dualprogram}
\end{equation}
The dual variable $Z=Z^\dagger$ is subject to $m$ equality constraints, defined by the $F_i$s and $c_i$s specified in the primal program, in addition to a condition of non-negativity, $Z\geq 0$. If there is a $Z>0$ satisfying the dual constraints and a $\vec{x}$ such that $F(\vec{x})>0$, the primal and dual problems are called strictly feasible. Under these conditions, the optimal primal and dual objective values are equal, the sets of optimal variables are non-empty, and furthermore, $F(\vec{x}_\text{opt})Z_\text{opt}=0$.

\section{Optimal LSD via symmetric extensions}\label{sec:LSDviasymm}
The main difficulty in the optimal LSD problem is that one does not know how to properly characterize the set of separable states. As already mentioned, for $\mathcal{H}_A\otimes\mathcal{H}_B=\mathbb{C}^2\otimes\mathbb{C}^2$ or $\mathbb{C}^3\otimes\mathbb{C}^2$, the separable states are exactly those that remain positive under partial transposition (PPT) \cite{peres96, horodecki96}, so that the positivity of $\rho_\text{sep}$ and its partial transpose $\rho_\text{sep}^{\text{T}_B}$ suffice to ensure separability. In these cases, finding the optimal LSD amounts precisely to solving a semidefinite program \cite{thiang09}. In this section, we will describe how to treat the general case of $\rho\in\mathcal{B}(\mathcal{H}_A\otimes\mathcal{H}_B)$.

\subsection{Symmetric extensions criterion for separability}
The existence of bound entangled states in higher dimensions prevents us from directly generalizing the above method. Nevertheless, Doherty, Parrilo, and Spedalieri have established a separability criterion (the DPS criterion) that is related to positivity \cite{doherty02, doherty04}. Specifically, a positive linear operator on $\mathcal{H}_A\otimes\mathcal{H}_B$ is separable iff it admits a Bose symmetric extension to any number of copies of $\mathcal{H}_B$. A $k$-Bose symmetric extension ($k$-BSE) of $\rho$ is a positive operator $\bar{\rho}\in\mathcal{B}(\mathcal{H}_A\otimes\mathcal{H}_B^{\otimes k})$ such that $\text{tr}_{B^{k-1}}\{\bar{\rho}\}=\rho$, and $\bar{\rho}$ is Bose symmetric, i.e., $\bar{\rho}(\mathbb{I}_A\otimes\Pi_\text{symm}^k)=\bar{\rho}$, where $\Pi_\text{symm}^k$ denotes the projector onto the symmetric subspace of $\mathcal{H}_B^{\otimes k}$. If in addition, $\bar{\rho}$ is PPT with respect to some bipartition $AB^l|B^{k-l}$, we call $\bar{\rho}$ a PPT-BSE of $\rho$. For convenience, we will only consider the partition $AB^{\lceil{k/2}\rceil}|B^{\lfloor{k/2}\rfloor}$, and define partial transposition to be transposition in the last $\lfloor {k/2} \rfloor$ copies of $\mathcal{H}_B$. A linear operator is also separable iff it admits a $k$-PPT-BSE for all natural numbers $k$, and we shall call this the DPS-PPT criterion. For brevity, $S_{(p)}^k$ will be used when making statements about $S^k$ and $S_p^k$ concurrently, and similarly for expressions such as DPS-(PPT).

The DPS-(PPT) criterion allows one to describe the set of separable states with a countably infinite set of positivity conditions. Let us define $S^k$ and $S_p^k$ to be the sets of (unnormalized) states which admit a $k$-BSE and $k$-PPT-BSE respectively, and let $S$ denote the convex cone of (unnormalized) separable states. Then the sets $\{S^k\}_{k=1}^\infty$ form a nested sequence of convex cones, i.e, $S^1\supseteq S^2\supseteq\ldots\supseteq S$, and similarly for $\{S_p^k\}_{k=1}^\infty$. This follows from the fact that a state with a $(k+1)$-(PPT)-BSE necessarily has a $k$-(PPT)-BSE, obtained by tracing over one copy of $\mathcal{H}_B$. Furthermore, $\lim_{k\rightarrow\infty} S_{(p)}^k=\bigcap_{k=1}^\infty S_{(p)}^k=S$ \cite{doherty04}.

\subsection{Approximating $S$ from the outside}\label{DPS}
Now, for each $k\in\mathbb{N}$, one can ask for the $k$-(PPT)-optimal decomposition $\rho=\lambda_k\varrho_k+(1-\lambda_k)\varsigma_k\equiv\tilde{\varrho}_k+\tilde{\varsigma}_k$ into a convex sum of a $k$-(PPT)-Bose symmetric extendible state and a positive remainder, in which the weight of the former is maximal. This is a convex optimization problem over the set $S_{(p)}^k$, with positivity constraints, and is in fact a semidefinite program (see Appendix \ref{explicitsdp} for details). We then have a \emph{sequence} of SDPs (of increasing size), indexed by $k$, where in each SDP, the objective function to be maximized is identical, while the feasible sets $\{S_{(p)}^k\}_{k=1}^\infty$ converge to $S$. Thus, even though the problem of finding the optimal LSD of an arbitrary state is not a semidefinite program \emph{per se}, and the feasible set $S$ is furthermore difficult to sample directly, the DPS criterion allows one to contruct a hierarchy of SDPs that ``approximate" the actual optimal LSD problem. Specifically, the sequence of $k$-(PPT)-optimal decompositions converges to the true optimal LSD. We shall now prove this statement.

Consider the sequence of $k$-optimal decompositions
\begin{align}
	\rho&=\lambda_1 \varrho_1 + (1-\lambda_1)\varsigma_1\nonumber\\
	    &=\lambda_2 \varrho_2 + (1-\lambda_2)\varsigma_2\nonumber\\
	    &\qquad\qquad\;\vdots\nonumber\\
	    &=\lambda_k \varrho_k + (1-\lambda_k)\varsigma_k\nonumber\\
	    &\qquad\qquad\;\vdots
\end{align}
which, \emph{a priori}, is not known to converge. Also, each $k$-optimal decomposition might not be unique; we will pick any optimal one as a representative. Observe that the sequence $\{\lambda_k\}_{k=1}^\infty$ is monotonically decreasing, since the $\lambda_k$s are the optimal values of the same objective function on decreasing feasible sets, and are bounded from below by the true degree of separability $\mathcal{S}$. Therefore, $\lambda_k\rightarrow\lambda_\infty\geq\mathcal{S}$ by the monotone convergence theorem. On the other hand, the sequence $\{\varrho_k\}_{k=1}^\infty$ might not be convergent, but it is contained in the compact set of normalized states, which has the Bolzano-Weierstrass property. That is, there is at least a subsequence $\{\varrho_{k_n}\}_{n=1}^\infty$ that is convergent (to some state $\varrho_\infty$). The limiting state of such a subsequence must be separable, because the terms in the subsequence get arbitrarily close to the closed set $S$. Therefore, the subsequence of $k$-optimal decompositions
\begin{align}
	\rho&=\lambda_{k_1} \varrho_{k_1} + (1-\lambda_{k_1})\varsigma_{k_1}\nonumber\\
	    &=\lambda_{k_2} \varrho_{k_2} + (1-\lambda_{k_2})\varsigma_{k_2}\nonumber\\
	    &\qquad\qquad\;\vdots\nonumber\\
	    &=\lambda_{k_n} \varrho_{k_n} + (1-\lambda_{k_n})\varsigma_{k_n}\nonumber\\
	    &\qquad\qquad\;\vdots\nonumber\\
	    &=\lambda_\infty\varrho_\infty + (1-\lambda_\infty)\varsigma_\infty\label{subdecomposition}
\end{align} 
converges to a valid LSD. By the definition of $\mathcal{S}$, $\lambda_\infty\leq\mathcal{S}$, but we already had $\mathcal{S}\leq\lambda_\infty$ earlier on, so that equality must hold. Since the optimal LSD is unique, the limiting decomposition in \eqref{subdecomposition} is in fact the optimal LSD, i.e., $\varrho_\infty=\varrho_\text{sep}$ and $\varsigma_\infty=\varsigma_\text{ent}$.
Note that the argument above holds for \emph{any} convergent subsequence. Now, if a sequence in a compact set has the property that every convergent subsequence has the same limit, then the sequence itself converges to that same limit. The proof of this assertion is by contradiction. Suppose $\varrho_k\nrightarrow\varrho_\text{sep}$. Then there exists an $\epsilon > 0$ and a subsequence of states $\{\varrho_{k'_n}\}_{n=1}^\infty$ such that $\norm{\varrho_{k'_n}-\varrho_\text{sep}}\geq\epsilon$ for all $n\in\mathbb{N}$. On the other hand, this subsequence is still contained in a compact set, and thus has its own convergent subsequence which, by hypothesis, should converge to $\varrho_\text{sep}$, in contradiction to $\varrho_k\nrightarrow\varrho_\text{sep}$. Hence, we must have $\varrho_k\rightarrow\varrho_\text{sep}$.

\begin{figure}
	\includegraphics[width=\linewidth,clip=]{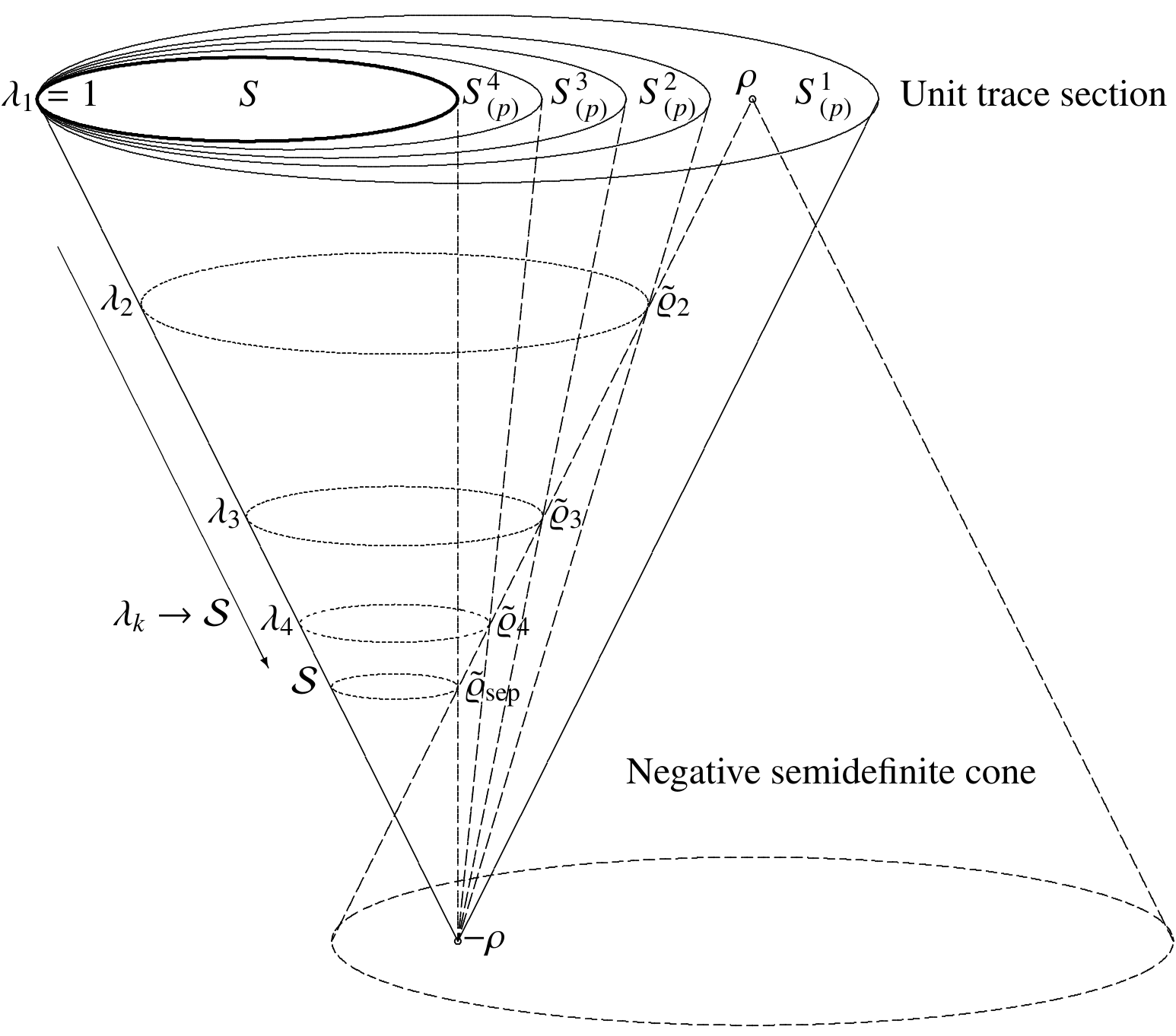}
\caption{Geometry of the approximation scheme using the DPS-(PPT) criterion. The feasible set in the $k$-th problem is the intersection of the affine cone $S^k_{(p)}-\rho$ with the negative semidefinite cone; among these candidates, $\tilde{\varrho}_k$ is the operator with the largest trace (highest position in the diagram). Here, $\rho$ lies inside $S^1_{(p)}\backslash S_p^2$, so that the first SDP has an optimal objective value $\lambda_1=1$, whereas the second SDP gives $\lambda_2 < 1$.}

\label{fig:cones}
\end{figure}

Therefore, the sequence of SDPs described above is a ``good" approximation to the actual optimal LSD problem in the sense that the optimal LSD can be provably obtained as the limit of the sequence of $k$-optimal decompositions. Note that the same argument holds for the $k$-PPT-optimal decompositions. One thus has a systematic numerical scheme to compute the optimal LSD of an arbitrary bipartite state, with the precision limited only by the computational resources available. In practice, each $\lambda_k$ provides an upper bound (of increasing tightness) for $\mathcal{S}$. Ideally, one would like to have lower bounds as well, which would be possible if we had a complementary scheme to approximate $S$ from the inside. Such a scheme exists: the so-called DPS* criterion, which was given by Navascu\'es, Owari, and Plenio \cite{navascues09, navascues09-2}.

\subsection{Approximating $S$ from the inside}\label{DPS*}
Navascu\'es \emph{et al.} \cite{navascues09-2} showed that the following convex cones, formed by a suitable perturbation of the sets $S_{(p)}^k$,
\begin{align}
	S^{*k}&\equiv\left\{\frac{k}{k+d_B}\sigma+\frac{d_B}{k+d_B}\text{tr}_B\{\sigma\}\otimes\frac{\mathbb{I}_B}{d_B}:\sigma\in S^k \right\},\nonumber\\
	S_p^{*k}&\equiv\left\{(1-\epsilon_k)\sigma+\epsilon_k\text{tr}_B\{\sigma\}\otimes\frac{\mathbb{I}_B}{d_B}:\sigma\in S_p^k \right\},\label{tildeS}
\end{align}
satisfy $S_{(p)}^{*k}\subseteq S$ for all $k$, and $\overline{\lim S_{(p)}^{*k}}=S$. Here, $d_B=\text{dim}(\mathcal{H}_B)$, $\mathbb{I}_B$ is the identity operator on $\mathcal{H}_B$, and $\epsilon_k$ is defined as
\begin{equation}
	\epsilon_k\equiv\frac{d_B}{2(d_B-1)}\min\left\{1-x:P_{\lfloor k/2 \rfloor+1}^{(d_B-2,\,k\:\text{mod}\:2)}(x)=0\right\},
\end{equation}
where $P_n^{(\alpha,\,\beta)}$ is a Jacobi polynomial \cite{abramowitz72}.
In other words, the sequence of sets $\{S_{(p)}^{*k}\}_{k=1}^\infty$ approximates $S$ from the inside, with (the closure of) the limiting set precisely equal to the set of separable states. Note, however, that we no longer have a hierarchy here --- the sequence $\{S_{(p)}^{*k}\}_{k=1}^\infty$ is not an increasing sequence of sets. Nevertheless, we can still construct a sequence of SDPs to approximate the optimal LSD problem, where this time, optimization is carried out over the sets $S_{(p)}^{*k}$. One then obtains a sequence of $k^*$-optimal decompositions
\begin{align}
	\rho&=\lambda^*_1 \varrho^*_1 + (1-\lambda^*_1)\varsigma^*_1\nonumber\\
	    &=\lambda^*_2 \varrho^*_2 + (1-\lambda^*_2)\varsigma^*_2\nonumber\\
	    &\qquad\qquad\;\vdots\nonumber\\
	    &=\lambda^*_k \varrho^*_k + (1-\lambda^*_k)\varsigma^*_k\nonumber\\
	    &\qquad\qquad\;\vdots
\end{align}
where each state $\varrho^*_k\in S_{(p)}^{*k}$. In this case, the sequence $\{\lambda_k^*\}_{k=1}^\infty$ is not necessarily monotonically increasing, but each $\lambda_k^*$ does provide a lower bound for $\mathcal{S}$.

As before, we would like to prove that this sequence of decompositions converges to the optimal LSD. This is in fact the case for any full-rank state $\rho$. The proof requires a different argument from the one used for the $k$-optimal decompositions. Now, for a full-rank state $\rho$, there is a strictly positive minimum eigenvalue $\mu_\text{min}$, as well as a degree of separability $\mathcal{S}\geq\mu_\text{min} > 0$ (since $\rho-\mu_\text{min}\gamma\geq 0$ for all normalized states $\gamma$, and in particular, for $\varrho_\text{sep}$). Then for any $\epsilon > 0$, after adding a traceless $\delta\varrho$ to $\varrho_\text{sep}$, the remainder

\begin{align}
	\rho-(\mathcal{S}-\epsilon)(\varrho_\text{sep}+\delta\varrho)&=\frac{\mathcal{S}-\epsilon}{\mathcal{S}}\left(\rho-\mathcal{S}\varrho_\text{sep}\right)\nonumber\\
	&\quad+\frac{\epsilon}{\mathcal{S}}\left[\rho-\frac{\mathcal{S}(\mathcal{S}-\epsilon)}{\epsilon}\delta\varrho\right]\nonumber\\
	&=\tilde{\varsigma}_\text{ent}+\frac{\epsilon}{\mathcal{S}}\left[\rho-\frac{\mathcal{S}(\mathcal{S}-\epsilon)}{\epsilon}\delta\varrho\right]\label{sqbracket}
\end{align}
is positive if the term in square brackets is positive. For instance, whenever $\norm{\delta\varrho}_\text{op}<\epsilon\mu_\text{min}$, where $\norm{\cdot}_\text{op}$ is the operator norm
\begin{equation}
	\norm{A}_\text{op}=\text{sup}\left\{\norm{Av}:\norm{v}\leq1\right\},
\end{equation}
we have
\begin{align}
\rho-\frac{\mathcal{S}(\mathcal{S}-\epsilon)}{\epsilon}\delta\varrho&\geq\rho-\frac{\mathcal{S}(\mathcal{S}-\epsilon)}{\epsilon}(\epsilon\mu_\text{min}\mathbb{I}_{AB})\nonumber\\
	&\geq\rho-\mu_\text{min}\mathbb{I}_{AB}\nonumber\\
	&\geq 0.
\end{align}
Here, $\mathbb{I}_{AB}$ is the identity operator on $\mathcal{H}_A\otimes\mathcal{H}_B$. Since $\overline{S_{(p)}^{*k}}\rightarrow S$, for all sufficiently large $k$, there will be some $\rho^*_k\in S_{(p)}^{*k}$ that is contained in the $\epsilon\mu_\text{min}$-neighbourhood of $\varrho_\text{sep}$. For these $\rho^*_k=\varrho_\text{sep}+\delta\varrho_k$, the inequality $\norm{\delta\varrho_k}_\text{op}<\epsilon\mu_\text{min}$ holds, so that the left-hand side of \eqref{sqbracket} is positive. Every such $\rho^*_k$ therefore appears in a $k^*$-suboptimal decomposition
\begin{equation}
	\rho=(\mathcal{S}-\epsilon)\rho_k^*+\tilde{\sigma}^*_{k,\text{remainder}}\;,
\end{equation}
which means that the $k^*$-optimal weight $\lambda^*_k$ must obey $\mathcal{S}-\epsilon\leq\lambda^*_k\leq\mathcal{S}$ for all large $k$. In other words, we have $\lambda^*_k\rightarrow\mathcal{S}$. By repeating the earlier argument for $\{\varrho_k\}_{k=1}^\infty$, one can easily show that every convergent subsequence of $\{\varrho_k^*\}_{k=1}^\infty$ converges to $\varrho_\text{sep}$, from which it follows that the entire sequence of $k^*$-optimal decompositions converges to the optimal LSD.

The argument above fails if $\rho$ does not have full rank, because then, the term in square brackets in \eqref{sqbracket} is not decidedly positive. One could of course restrict the choice of $\delta\rho$ to lie in the support of $\rho$, but with this restriction, $\varrho_\text{sep}+\delta\rho$ might not be found inside $S_{(p)}^{*k}$ for any $k$. As an extreme example, consider a separable $\rho$ that has a rank smaller than $d_B$. An inspection of \eqref{tildeS} reveals that $S_{(p)}^{*k}$ contains only states with ranks that are least $d_B$. As a result, the only permissible $k^*$-decomposition of $\rho$ is the trivial one, i.e., $\lambda^*_k=0$ for all $k$. On the other hand, the true degree of separability is $\mathcal{S}=1$, so we will fail to obtain the optimal LSD by taking the limit of the trivial $k^*$-optimal decompositions (see also Fig.~\ref{bound4x2plot}).

\subsection{Numerical results}
\begin{figure}[h]
	\includegraphics[width=\linewidth,clip=]{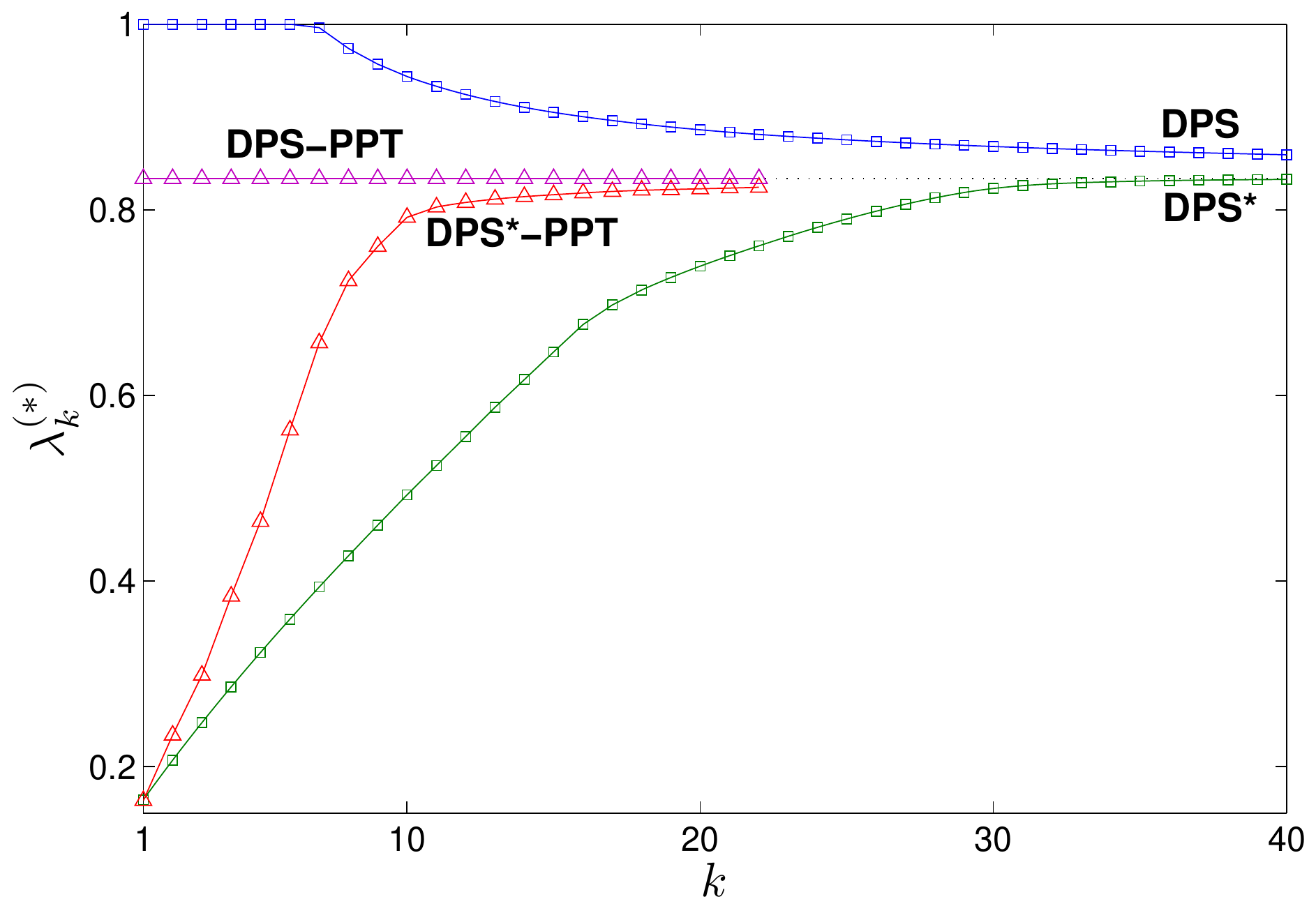}
		\caption{Optimal objective values for a generic full-rank state in $\mathcal{B}(\mathbb{C}^2\otimes\mathbb{C}^2)$, using both the DPS-(PPT) and DPS*-(PPT) criteria. The dotted line indicates the best upper bound for $\mathcal{S}$, which is exact in this case. The best lower bound is provided by the DPS* criterion, and is within $10^{-3}$ of the best upper bound.}
		\label{generic2x2plot}
\end{figure}

\begin{figure}
	\includegraphics[width=\linewidth,clip=]{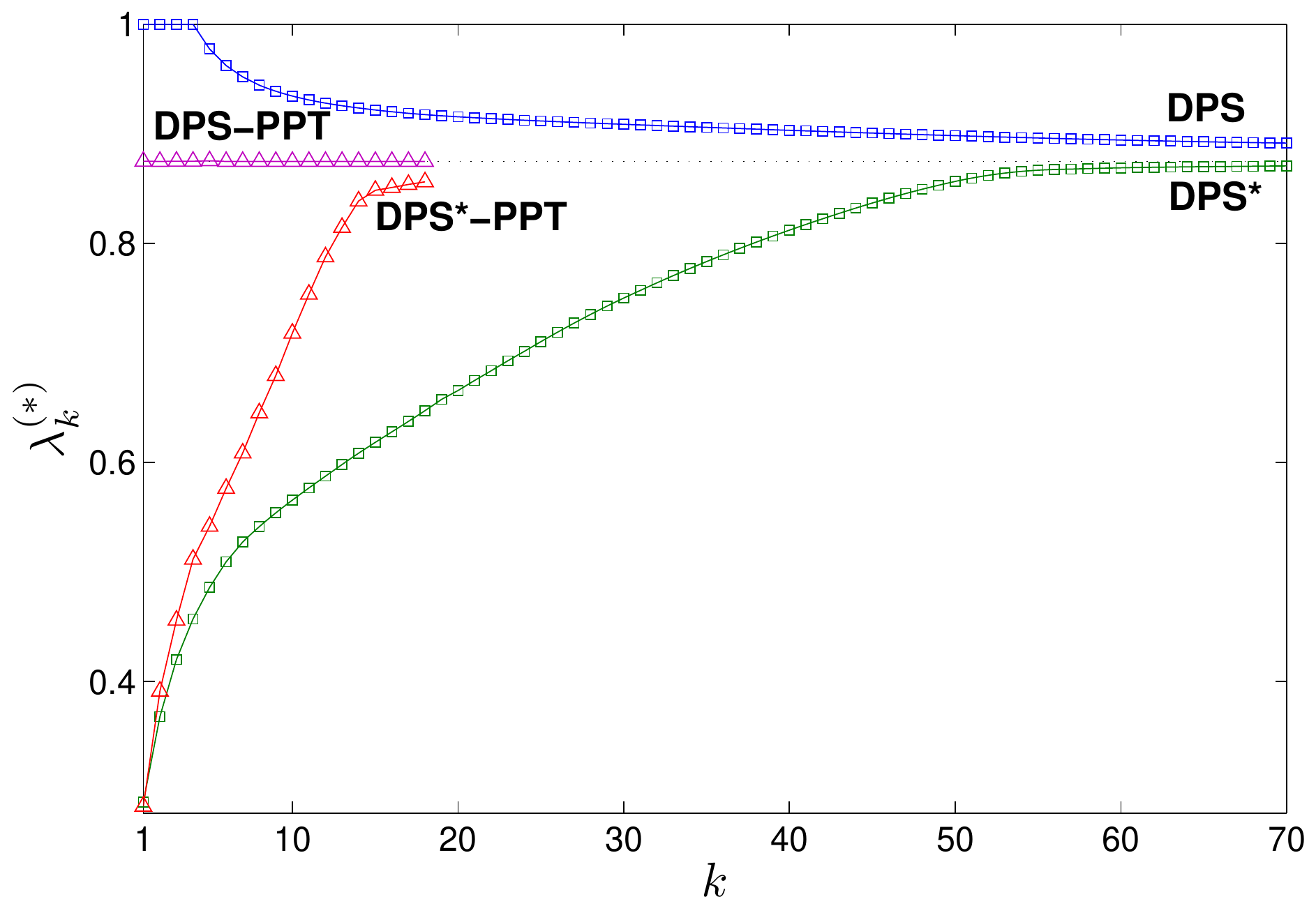}
		\caption{Optimal objective values for a generic full-rank state in $\mathcal{B}(\mathbb{C}^4\otimes\mathbb{C}^2)$. The PPT criterion already provides what appears to be the true degree of separability. As in the $\mathbb{C}^2\otimes\mathbb{C}^2$ case, the difference between the best upper and lower bounds for $\mathcal{S}$ is very small.}
		\label{generic4x2plot}
\end{figure}

\begin{figure}
	\includegraphics[width=\linewidth,clip=]{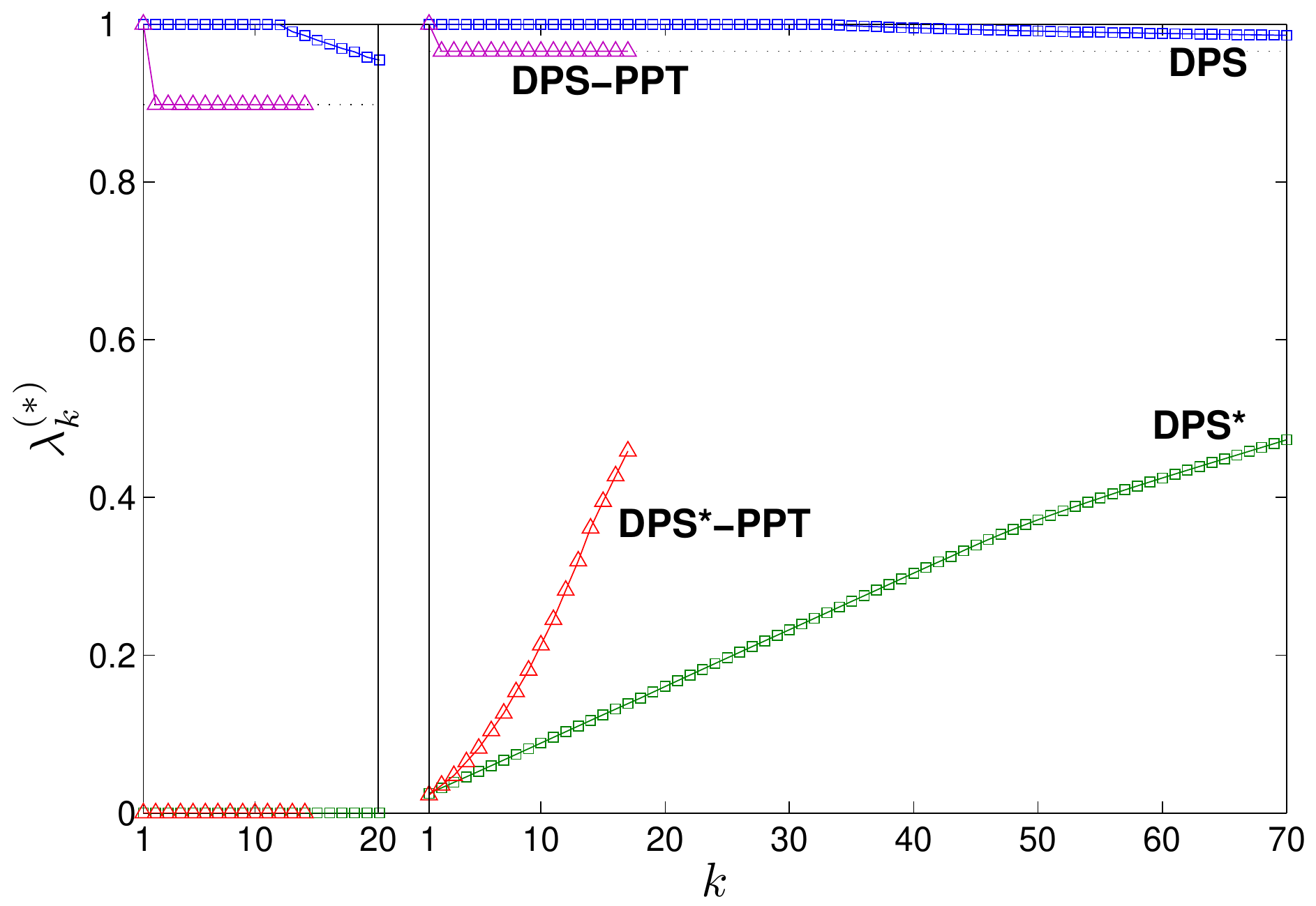}
		\caption{Optimal objective values for a bound entangled state in $\mathcal{B}(\mathbb{C}^4\otimes\mathbb{C}^2)$ (left). $\lambda_1=1$ with the DPS-PPT criterion reflects that $\rho$ is PPT. The second test with $\lambda_2 < 1$ then reveals the (bound) entanglement of $\rho$. As $\rho$ has reduced-rank, the optimal objective values obtained from the DPS*-(PPT) criteria are all zero, and fail to converge to $\mathcal{S}$. After mixing a small amount of the identity operator with $\rho$ (right), non-trivial $k^*$-optimal decompositions are obtained, and $\lambda^*_k$ begins to climb towards $\mathcal{S}$.}
		\label{bound4x2plot}
\end{figure}

\begin{figure}
	\includegraphics[width=\linewidth,clip=]{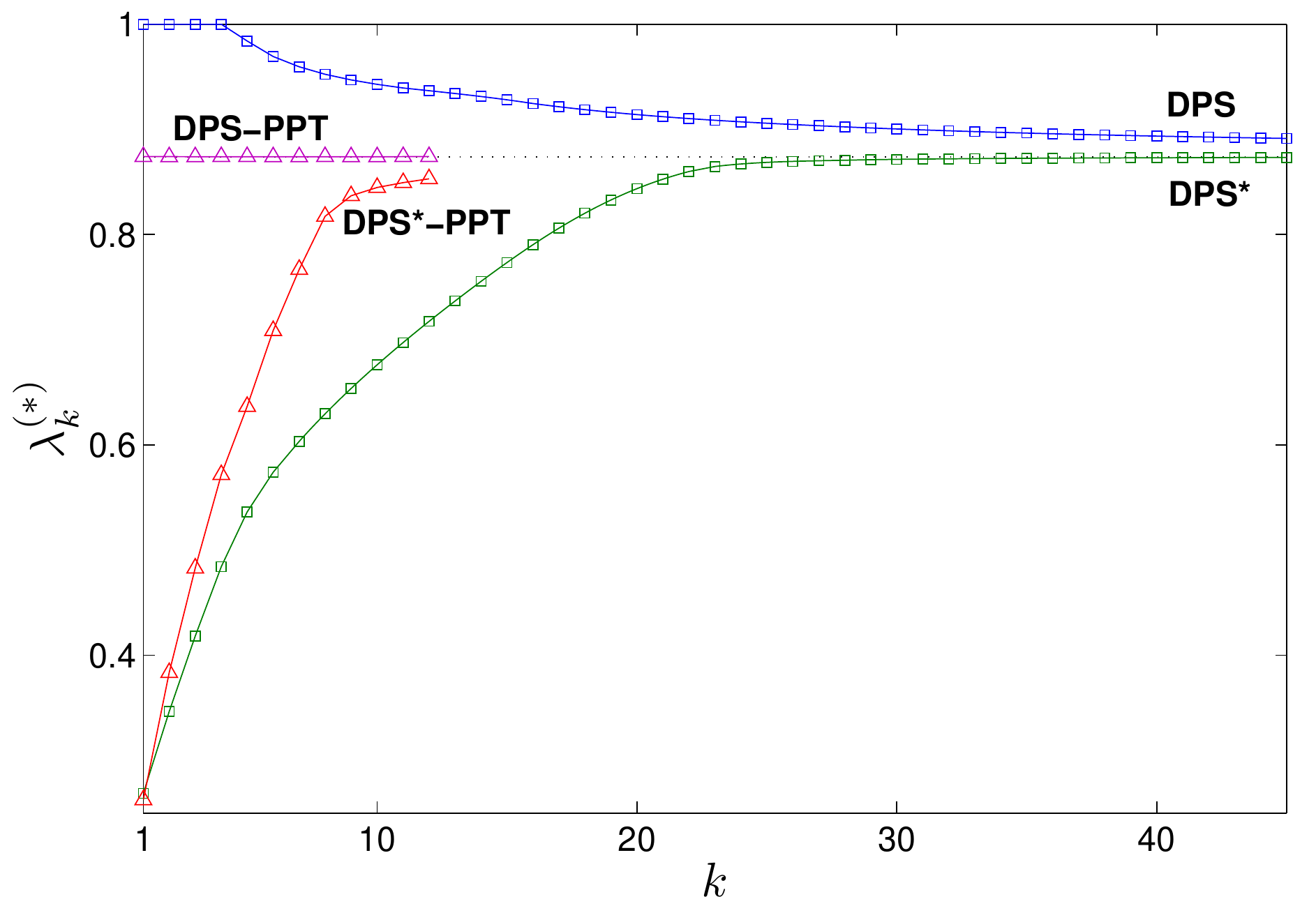}
		\caption{Optimal objective values for a generic full-rank state in $\mathcal{B}(\mathbb{C}^5\otimes\mathbb{C}^2)$. It appears that good upper and lower bounds for $\mathcal{S}$ can be obtained as long as qubit extensions are being considered. The DPS-PPT criterion performs best.}
		\label{generic5x2plot}
\end{figure}

\begin{figure}
	\includegraphics[width=\linewidth,clip=]{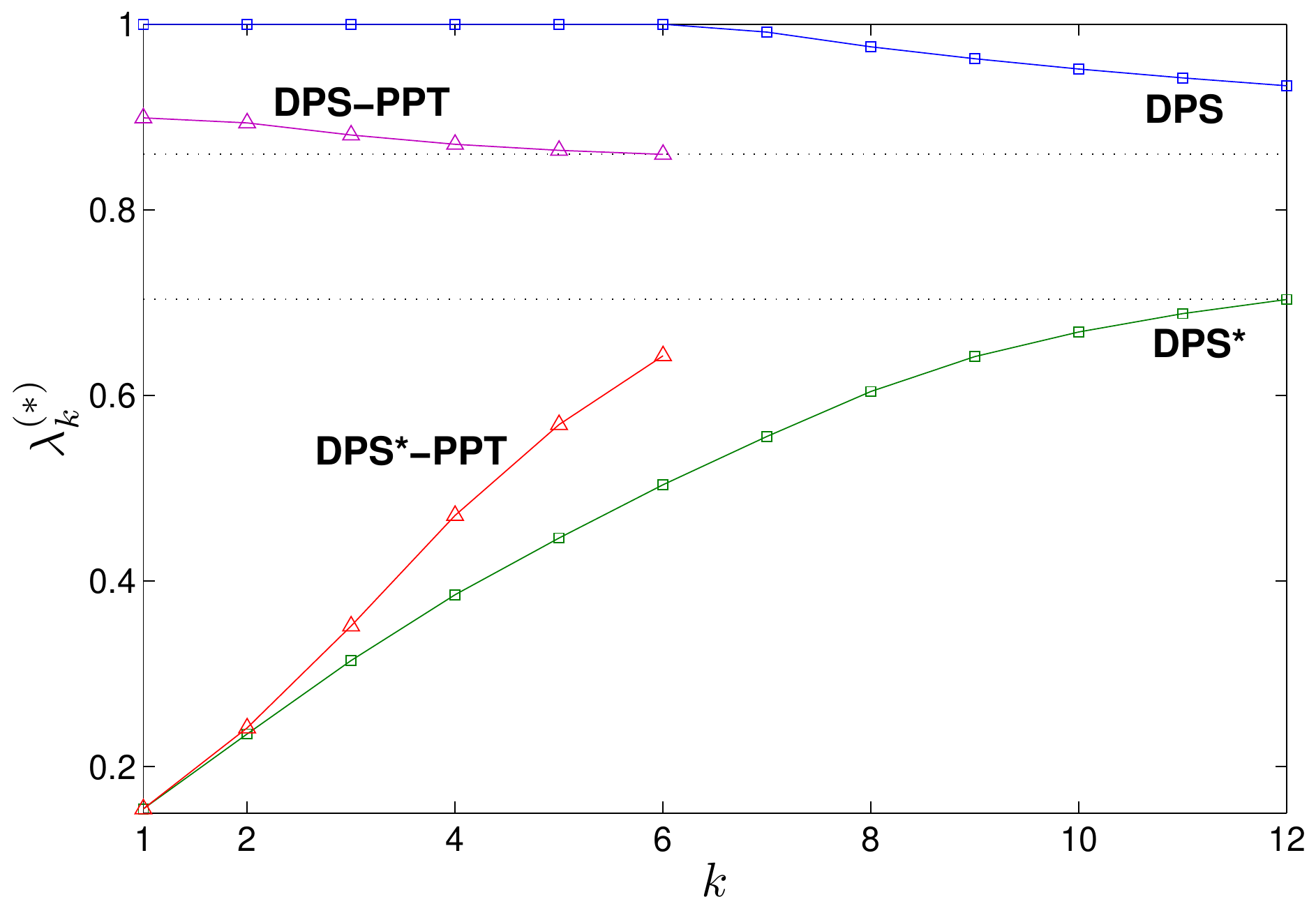}
		\caption{Optimal objective values for a generic full-rank state in $\mathcal{B}(\mathbb{C}^3\otimes\mathbb{C}^3)$. The maximum number of qutrit extensions that can be considered is limited by the available memory. A sizable gap between the best bounds remains after 12 extensions were used. This gap is comparable to the gap for qubit extensions at the same stage. More significantly, the $\lambda_k$ values obtained from the DPS-PPT criterion does not immediately flatten out after a few extensions.}
		\label{generic3x3plot}
\end{figure}

For our numerical work, we introduce a minor modification to the definition of $S_p^1$: for partial transposition, we use the partition $A|B$ for consistency with the usual notion of PPT in bipartite states. $S_p^1$ is thus the set of PPT states. We first test out the formulation described above using a full-rank two-qubit state (Fig.~\ref{generic2x2plot}). Since there are no bound entangled states in this case ($S_p^1=S_p^2=\ldots=S$), the SDPs using the DPS-PPT criterion give the exact degree of separability from the very first level in the hierarchy. The lower bounds provided by the DPS* and DPS*-PPT criteria are seen to converge to $\mathcal{S}$ as the number of extensions of $\mathcal{H}_B$ increases, with the best lower bound within 0.001 of $\mathcal{S}$. Although the convergence with the DPS criterion is much slower than with the DPS-PPT criterion, we do obtain some useful information, namely that $\rho$ lies in $S^6\backslash S^7$, since $\lambda_6=1, \lambda_7 < 1$.

Next, we consider the smallest system in which the PPT separability criterion fails. A generic full-rank ququart-qubit state is used in Fig.~\ref{generic4x2plot}. The additional PPT constraint is computationally more expensive, but convergence is again much faster (with respect to the number of extensions used). The PPT criterion with a single copy of $\mathcal{H}_B$ already attains what appears to be the true degree of separability. Also, the gap between the best upper bound and best lower bound is less than $0.005$.

We are also interested in what happens for a bound entangled state. In Fig.~\ref{bound4x2plot}, the state $\rho$ used for the plot on the left is the bound entangled ququart-qubit state from the one-parameter family introduced in \cite{horodecki97}, with $b=0.5$. Note that the first PPT test gives an optimal objective value of $1$, consistent with the fact that $\rho$ is PPT-entangled, lying in $S_p^1\backslash S_p^2$. The optimal objective values obtained from the DPS* approximation fail to converge, because $\rho$ has reduced-rank (see Section \ref{DPS*}). After mixing a small amount of the identity operator with $\rho$, one still has a bound entangled state, which is now full-rank. Then, the DPS* approximation works, albeit rather slowly, as is evident from the plot on the right of Fig.~\ref{bound4x2plot}.

The procedure continues to perform well for $\rho\in\mathcal{B}(\mathbb{C}^5\otimes\mathbb{C}^2)$ (Fig.~\ref{generic5x2plot}), where the gap between the best upper bound and best lower bound for $\mathcal{S}$ is less than 0.001. When we move from qubit extensions to qutrit extensions, the dimension of the SDP grows faster with $k$, with computational resources limiting the number of extensions that can be considered. For instance, in Fig.~\ref{generic3x3plot}, there is still a gap of about 0.15 between the best upper bound and the best lower bound after using up to twelve qutrit extensions.

We remark that in our calculations, memory issues prevent us from computing the $k$-PPT-optimal decompositions beyond a certain value of $k$. This is reflected in the early termination of the PPT curves in Figs.~\ref{generic2x2plot} to \ref{generic3x3plot}, and will be discussed in the next subsection.

\subsection{Complexity considerations}
At first glance, the size of the semidefinite program seems to increase at least exponentially with the number of extensions $k$ used. After all, the dimension of $\mathcal{H}_A\otimes\mathcal{H}_B^{\otimes k}$ is $d_Ad_B^k$, which is exponential in $k$, and we are presumably optimizing over $\mathcal{B}(\mathcal{H}_A\otimes\mathcal{H}_B^{\otimes k})$. However, it actually suffices to consider only the subset $\mathcal{B}(\mathcal{H}_A\otimes\mathcal{H}_{B, \text{symm}}^{\otimes k})$ of $\mathcal{B}(\mathcal{H}_A\otimes\mathcal{H}_B^{\otimes k})$, namely, the positive linear operators acting on the symmetric (with respect to interchange of $\mathcal{H}_B$) subspace $\mathcal{H}_A\otimes\mathcal{H}_{B, \text{symm}}^{\otimes k}$. The dimension of this subspace is
\begin{align}
	\text{dim}\left(\mathcal{H}_A\otimes\mathcal{H}_{B, \text{symm}}^{\otimes k}\right)&=d_A\begin{pmatrix} d_B+k-1\\k \end{pmatrix}\nonumber\\
	&=\frac{d_A(k+1)\ldots(d_B+k-1)}{(d_B-1)!}\nonumber\\
	&\leq\frac{d_A(d_B+k-1)^{d_B-1}}{(d_B-1)!},
\end{align}
which is at most polynomial in $k$. In particular, when $d_B=2$, the growth in dimension is linear, which greatly facilitates the large-$k$ calculations for qudit-qubit states. The number of real variables needed to parameterize a candidate $\bar{\rho}_k$ is $\text{dim}(\mathcal{H}_A\otimes\mathcal{H}_{B, \text{symm}}^{\otimes k})^2$. The numerical algorithms used for SDP problems typically involve solving least-squares problems, each requiring $O(m^2n^2)$ steps, where $m$ is the number of variables and $n$ is the size of the matrices involved in the problem. The number of iterations required scales no worse than $O(n^{1/2})$. Therefore, for fixed $d_A, d_B$, the complexity of the SDPs with the DPS criterion scales polynomially with $k$.

If one wishes to consider PPT-symmetric extensions, it suffices to consider positive operators on a subspace of $\mathcal{H}_A\otimes\mathcal{H}_B^{\otimes k}$ that is isomorphic to $\mathcal{H}_A\otimes\mathcal{H}_{B, \text{symm}}^{\otimes {\lceil k/2 \rceil}}\otimes\mathcal{H}_{B, \text{symm}}^{\otimes {\lfloor k/2 \rfloor}}$ \cite{doherty04}. Imposing an additional PPT criterion thus introduces an additional matrix block of size $n'\times n'$, where
\begin{equation}
	n'=d_A\begin{pmatrix}d_B+\lceil k/2 \rceil-1\\\lceil k/2 \rceil\end{pmatrix}\begin{pmatrix}d_B+\lfloor k/2 \rfloor-1\\\lfloor k/2 \rfloor\end{pmatrix},
\end{equation}
so the complexity with the DPS-PPT criterion remains polynomial. However, the memory resources needed to handle the larger matrices when using the DPS-PPT criterion is generally much greater than that required when the PPT constraint is dropped. When memory is a limiting factor, the DPS* criterion allows more extensions to be accessed, which may sometimes lead to better lower bounds than those obtained using the DPS*-PPT criterion with fewer extensions (see Figs.~\ref{generic2x2plot} to \ref{generic3x3plot}).

\section{Optimal LSD of low-rank states}\label{sec:lowrank}
In this section, we will investigate the optimal LSDs of some low-rank states. The first subsection deals with two-qubit states, and is followed by a straightforward generalization to qubit-qutrit states. Rank-2 states in any dimension are then considered in the last subsection.

\subsection{Two-qubit states}
It was also noted \cite{doherty04} that the entire sequence of tests is required to fully characterize $S$ for $d_A \times d_B > 6$. That is, for every $k$, there exist entangled states lying in $S_{(p)}^k$ but not in $S_{(p)}^{k+1}$. Furthermore, for every $k$, the volume of the set of such states, as quantified by the measure introduced in \cite{zyczkowski98}, is non-zero. Only when $d_A \times d_B \leq 6$ does a $k$th PPT test suffice (the first one). In these cases, a more detailed study of the optimal LSDs is possible. For two qubits, analytical expressions for the optimal LSDs were found for some special classes of states \cite{englert00}, while an algebraic way for handling the remaining states was described in \cite{wellens01}. A unifying approach using semidefinite programming was recently reported in \cite{thiang09}. The following is essentially a summary of that work, presented here to illustrate the use of SDP in the optimal LSD problem.

One of the salient features of semidefinite programming is its well-developed duality theory, which one can exploit to extract valuable information about the problem at hand. For example, one can obtain optimality conditions using the so-called complementary slackness condition. This says that under the conditions of strict primal and dual feasibility, the optimal primal and dual variables have orthogonal ranges. 

In the optimal LSD problem for two-qubits, $-\mathcal{S}$ is found by minimizing the linear objective function $-\text{tr}\{\tilde{\rho}_\text{sep}\}$ subject to three positivity constraints, namely, $\tilde{\rho}_\text{sep}\geq 0, \tilde{\rho}_\text{sep}^{\text{T}_B}\geq 0$, and $\rho-\tilde{\rho}_\text{sep}\geq 0$, which we can write as a single matrix inequality
\begin{equation}
\begin{bmatrix}
		\tilde{\rho}_\text{sep} & 0 & 0\\
		0 & \tilde{\rho}_\text{sep}^{\text{T}_B} & 0\\
		0 & 0 & \rho-\tilde{\rho}_\text{sep}
\end{bmatrix}\geq 0.
\end{equation}
We parameterize $\tilde{\rho}_\text{sep}$ by its sixteen real components with respect to some basis for $\mathcal{B}(\mathbb{C}^2\otimes\mathbb{C}^2)$. For example, we can use the basis $\{\sigma_i\otimes\tau_j\}_{i,j=0}^3$, where $\sigma_0=\tau_0=\mathbb{I}_2$ and $\sigma_i, \tau_j, i,j=1,2,3$ are the Pauli matrices. Writing these basis elements as $\{E_i\}_{i=1}^{16}$, and defining $F_i=\text{diag}(E_i, E_i^{\text{T}_B}, -E_i)$ and $\vec{c}^\text{T}=(-1,0,\ldots,0)$, we obtain the primal SDP in the form \eqref{primalprogram}. As for the corresponding dual variable $Z$, we may also consider it to be block-diagonal: $Z=\text{diag}(Z_1, Z_2, Z_3)$. When $\rho$ is full-rank, one easily verifies that both the primal and dual problems are strictly feasible (choose $\tilde{\rho}_\text{sep}$ and $Z$ to be suitable multiples of the identity). Consequently, we have the necessary and sufficient condition for optimality:
\begin{equation}
	\begin{array}{ll}
	\text{(i)} & \begin{bmatrix}
		\tilde{\varrho}_\text{sep} & 0 & 0\\
		0 & \tilde{\varrho}_\text{sep}^{\,\text{T}_B} & 0\\
		0 & 0 & \tilde{\varsigma}_\text{ent}
		\end{bmatrix}
		\begin{bmatrix}
		\mathcal{Z}_1 & 0 & 0\\
		0 & \mathcal{Z}_2 & 0\\
		0 & 0 & \mathcal{Z}_3
		\end{bmatrix}=0,\\ \\
	\text{(ii)} & \begin{bmatrix}
		\tilde{\varrho}_\text{sep} & 0 & 0\\
		0 & \tilde{\varrho}_\text{sep}^{\,\text{T}_B} & 0\\
		0 & 0 & \tilde{\varsigma}_\text{ent}
		\end{bmatrix}\geq 0,\\ \\
	\text{(iii)} & \begin{bmatrix}
		\mathcal{Z}_1 & 0 & 0\\
		0 & \mathcal{Z}_2 & 0\\
		0 & 0 & \mathcal{Z}_3
		\end{bmatrix}\geq 0,\quad \text{tr}\{\tilde{F}_i\mathcal{Z}\}=c_i,\;i=1,\ldots,16.
	\end{array}\label{compslack}
\end{equation}

The dual constraints can be used to express $\mathcal{Z}_3$ in terms of $\mathcal{Z}_1$ and $\mathcal{Z}_2$, and after some algebra, we arrive at the optimality conditions
\begin{align}
	(\mathcal{Z}_1+\mathcal{Z}_2^{\text{T}_B})\tilde{\varsigma}_\text{ent}&=-\tilde{\varsigma}_\text{ent},\nonumber\\
	\tilde{\varrho}_\text{sep}^{\text{T}_B}\mathcal{Z}_2&=0.\label{SDPWK2}
\end{align}
These are identical to the equations originally found by Wellens and Ku\'s \cite{wellens01}, here derived using the duality theory of SDP. With an appropriate reparameterization (see \cite{thiang09}), the rank-3 case can be handled likewise, giving rise to another set of optimality conditions, hereafter termed the generalized Wellens--Ku\'s equations.

\subsection{Qubit-qutrit states}
The optimal LSD for qubit-qutrit states, can be found in an almost identical fashion. In fact, one needs only to select an appropriate basis, for example, by replacing $\{\tau_j\}_{j=0}^3$ with $\mathbb{I}_3$ along with the generators of SU$(3)$. For full-rank states, the optimal decomposition again satisfies a set of generalized Wellens--Ku\'s equations, which are identical to those for the two-qubit case, except that $\varrho_\text{sep}$ can now be rank-4 or rank-5 while $\varsigma_\text{ent}$ can be rank-1 or rank-2. Indeed, the entangled remainder in the optimal LSD can have rank up to $d_A+d_B-1$ \cite{lewenstein98, karnas01}. In analogy to the two-qubit case \cite{wellens01}, we remark here that when $\varrho_\text{sep}$ and $\varrho_\text{sep}^{\text{T}_B}$ have their highest allowed rank (6 and 5 respectively), and when $\varsigma_\text{ent}$ is a pure state, $\varsigma_\text{ent}$ is in fact maximally entangled. This can be seen as follows. Since $\varrho_\text{sep}$ has full-rank, $\mathcal{Z}_1$ vanishes by the complementary slackness condition. Meanwhile, $\mathcal{Z}_2$ must be rank-1 since its range is orthogonal to $\varrho_\text{sep}^{\text{T}_B}$, which has rank 5. Now, it can be shown that for a pure state $\gamma\in\mathcal{B}(\mathbb{C}^2\otimes\mathbb{C}^N)$ with concurrence $q$ \cite{wootters97, wootters98, rungta01} (normalized to take values in $[0,1]$), the partially transposed state $\gamma^{\text{T}_B}$ has eigenvalues $\frac{1\pm p}{2},\pm\frac{q}{2}$, and $0$, where $p\equiv\sqrt{1-q^2}$. In particular, the eigenstate associated with the non-degenerate negative eigenvalue is a maximally entangled state. Therefore, the first eigenvalue equation in \eqref{SDPWK2} with $\mathcal{Z}_1=0$ states that $\varsigma_\text{ent}$ is maximally entangled.

\subsection{Optimal LSD of rank-2 states in $\mathbb{C}^M\otimes\mathbb{C}^N$}
The optimal LSD for rank-2 two-qubit states was worked out in \cite{englert02}. Here, we describe a prescription to find the optimal LSD for rank-2 states in $\mathbb{C}^M\otimes\mathbb{C}^N$. No use is made of SDP, as the optimization can actually be carried out analytically. First, we recall some early definitions and theorems introduced in \cite{lewenstein98, karnas01}. 

\begin{definition}\label{maximallambda}
	A non-negative parameter $\Lambda$ is said to be maximal with respect to a positive (possibly unnormalized) operator $\tilde{\rho}$ and a projector $P=\ket{\psi}\bra{\psi}$ if $\tilde{\rho}-\Lambda P$ is positive, but $\tilde{\rho}-(\Lambda+\epsilon) P$ is not positive for every $\epsilon > 0$.
\end{definition}

\begin{definition}\label{maximalpair}
	A pair of non-negative parameter $(\Lambda_1,\Lambda_2)$ is said to be maximal with respect to a positive operator $\tilde{\rho}$ and a pair of projectors $(P_1=\ket{\psi_1}\bra{\psi_1}, P_2=\ket{\psi_2}\bra{\psi_2})$, if (i) $\tilde{\rho}-(\Lambda_1 P_1 + \Lambda_2 P_2)$ is positive, (ii) $\Lambda_1$ is maximal with respect to $\tilde{\rho}-\Lambda_2 P_2$ and $P_1$ while $\Lambda_2$ is maximal with respect to $\tilde{\rho}-\Lambda_1 P_1$ and $P_2$ (in the sense of Definition \ref{maximallambda}), (iii) the sum $\Lambda_1+\Lambda_2$ is maximal.
\end{definition}
Maximality as defined above is characterized by the following lemma, proven in \cite{lewenstein98}. Here, $\mathcal{R}(\rho)$ refers to the range of $\rho$, while $\rho^{-1}$ refers to the pseudo-inverse of $\rho$ if $\rho$ has reduced-rank.
\begin{lemma}\label{Lambdavalue}
	\singlespace
	The maximal $\Lambda$ with respect to $\rho$ and $\ket{\psi}\bra{\psi}$ is given by	
	\begin{itemize}
	\item[(a)]{$0$, if $\ket{\psi}\notin \mathcal{R}(\rho)$}.
	\item[(b)]{${\bra{\psi}\rho^{-1}\ket{\psi}}^{-1}$, if $\ket{\psi}\in \mathcal{R}(\rho)$}.
	\end{itemize}
	The maximal $(\Lambda_1, \Lambda_2)$ with respect to $\rho$ and $\left(\ket{\psi_1}\bra{\psi_1}, \ket{\psi_2}\bra{\psi_2}\right)$ is given by
	\begin{itemize}
		\item[(a)]{$\left(0,0\right)$, if $\ket{\psi_1}, \ket{\psi_2}\notin \mathcal{R}(\rho)$},
		\item[(b)]{$\left(0,\bra{\psi_2}\rho^{-1}\ket{\psi_2}^{-1}\right)$, if $\ket{\psi_1}\notin\mathcal{R}(\rho)$ and $\ket{\psi_2}\in\mathcal{R}(\rho)$},
		\item[\emph{(c)}]{$\left(\bra{\psi_1}\rho^{-1}\ket{\psi_1},\bra{\psi_2}\rho^{-1}\ket{\psi_2}\right)$, \\
		if $\ket{\psi_1}, \ket{\psi_2}\in \mathcal{R}(\rho)$ and $\bra{\psi_1}\rho^{-1}\ket{\psi_2}=0$},
		\item[(d)]{$\frac{1}{D}\Bigl(\bra{\psi_2}\rho^{-1}\ket{\psi_2}-|\bra{\psi_1}\rho^{-1}\ket{\psi_2}|, \bra{\psi_1}\rho^{-1}\ket{\psi_1}\\
		-|\bra{\psi_2}\rho^{-1}\ket{\psi_1}|\Bigr)$,\\
		with $D=\bra{\psi_1}\rho^{-1}\ket{\psi_1}\bra{\psi_2}\rho^{-1}\ket{\psi_2}-|\bra{\psi_1}\rho^{-1}\ket{\psi_2}|^2$, \\
		if $\ket{\psi_1}, \ket{\psi_2}\in \mathcal{R}(\rho)$ and $\bra{\psi_1}\rho^{-1}\ket{\psi_1}, \bra{\psi_2}\rho^{-1}\ket{\psi_2}\geq |\bra{\psi_1}\rho^{-1}\ket{\psi_2}|\neq 0$},
		\item[(e)]{$\left(\bra{\psi_1}\rho^{-1}\ket{\psi_1}^{-1},0\right)$, if $\ket{\psi_1}, \ket{\psi_2}\in \mathcal{R}(\rho)$ and $\bra{\psi_1}\rho^{-1}\ket{\psi_1}\geq|\bra{\psi_1}\rho^{-1}\ket{\psi_2}|\geq\bra{\psi_2}\rho^{-1}\ket{\psi_2}$}.
	\end{itemize}
\end{lemma}
The following theorem then characterizes the best separable approximation $\varrho_\text{sep}$ in the optimal LSD \cite{lewenstein98, lewenstein00}.
\begin{theorem}\label{BSAtheorem}
	Let $V$ be the set of all normalized product vectors $v_\alpha$ in the range of $\rho$, indexed by $\alpha$. Then $\tilde{\varrho}_\text{sep}=\sum_\alpha{\Lambda_\alpha P_\alpha}$ iff (i) each $\Lambda_\alpha$ is maximal with respect to $\rho-\sum_{\alpha'\neq\alpha}{\Lambda_{\alpha'} P_{\alpha'}}$ and $P_\alpha$, and (ii) every pair $(\Lambda_\alpha, \Lambda_\beta)$ is maximal with respect to $\rho-\sum_{\alpha'\neq\alpha,\,\beta}{\Lambda_{\alpha'} P_{\alpha'}}$ and $(P_\alpha, P_\beta)$.
\end{theorem}

Note that the set of product vectors does not constitute a vector space. Furthermore, the set $V$ is generally not finite (and indeed, even uncountable), so that direct computation of the optimal LSD using Theorem \ref{BSAtheorem} is very difficult. However, we will prove that for entangled rank-2 states, $V$ contains at most two terms. Then, if $V=\emptyset$, the optimal LSD is just the trivial decomposition. If $V=\{\ket{\psi}\}$, only condition (i) in Theorem \ref{BSAtheorem} is relevant. If $V=\{\ket{\psi_1}, \ket{\psi_2}\}$, there is only one pair $(\Lambda_1, \Lambda_2)$ to consider in condition (ii) of Theorem \ref{BSAtheorem}, which by Definition \ref{maximalpair} already implies condition (i). In any case, Lemma \ref{Lambdavalue} provides the optimal LSD as it gives explicitly the maximal $\Lambda_\alpha$ required in $\tilde{\varrho}_\text{sep}=\sum_\alpha{\Lambda_\alpha P_\alpha}$. It remains to prove the assertion that an entangled rank-2 state has at most two product vectors (up to normalization) in its range.

In the spectral decomposition of an entangled $\rho_\text{rank-2}=\mu_1\ket{\phi_1}\bra{\phi_1}+\mu_2\ket{\phi_2}\bra{\phi_2}$, at least one eigenvector, say $\ket{\phi_1}$ is entangled. Using the Schmidt basis for $\ket{\phi_1}$ and assuming that $M\leq N$, we can write the range of $\rho_\text{rank-2}$ as
\begin{equation}
	\mathcal{R}(\rho_\text{rank-2})=\text{span}\left\{\ket{\phi_1},\ket{\phi_2}\right\}=
	\text{span}\left\{\:\begin{pmatrix}\eta_1 \\ 0 \\ \vdots \\ \eta_k \\ \vdots \\ \eta_M \\ \vdots \\ 0 \end{pmatrix}\,,\,
																										\begin{pmatrix}z_{11} \\ z_{12}\\ \vdots \\ z_{kk} \\ \vdots \\ z_{MM} \\ \vdots \\ z_{MN}\end{pmatrix}\:
	\right\},
\end{equation}
where the Schmidt coefficients $\eta_k \geq 0$ appear (in decreasing order) in the $\left((k-1)N+k\right)$-th row, and $z_{ij}\in\mathbb{C}$. Now we look for product vectors in the range of $\rho_\text{rank-2}$, i.e., $\ket{\psi}=\chi\ket{\phi_1}+\ket{\phi_2}\;\chi\in\mathbb{C}$. To facilitate this search, we require a simple lemma.

\begin{lemma}\label{productlemma}
	A vector $\vec{a}^\text{T}=(a_{11}, \ldots, a_{1N}, a_{21}, \ldots, a_{MN})^\text{T}\in\mathbb{C}^M\otimes\mathbb{C}^N$, written in a product basis, is a product vector iff $a_{ij}a_{kl}=a_{il}a_{kj}$ for all $i,k=1,\ldots,M$ and $j,l=1,\ldots,N$.
\end{lemma}
Necessity is easy to prove, using
\begin{equation}
\begin{pmatrix}a_{11}\\\vdots\\a_{MN}\end{pmatrix}=\begin{pmatrix}b_1\\\vdots\\b_M\end{pmatrix}\otimes\begin{pmatrix}c_1\\\vdots\\c_N\end{pmatrix}=\begin{pmatrix}b_1c_1\\\vdots\\b_1c_N\\\vdots\\b_Mc_1\\\vdots\\b_Mc_N\end{pmatrix}.
\end{equation}
Now, suppose $a_{ij}a_{kl}=a_{il}a_{kj}$ for all $i,j,k,l$. Since $\vec{a}\neq\vec{0}$, there must be some $i_0, j_0$ such that $a_{i_0j_0}\neq 0$. Then $a_{ij_0}a_{kl}-a_{il}a_{kj_0}=0$ implies that
\begin{equation}
	\begin{pmatrix}a_{il}\\a_{kl}\end{pmatrix}=\nu_l^{ik}\begin{pmatrix}a_{ij_0}\\a_{kj_0}\end{pmatrix}\qquad\text{$\forall i,k,l$},\label{productcondition}
\end{equation}
for some set of complex numbers $\{\nu_l^{ik}\}$, except possibly when ${(a_{ij_0}, a_{kj_0})=(0,0)}$ but $(a_{il},a_{kl})\neq(0,0)$. But this exceptional case cannot happen, because if, for example, $a_{kl}\neq 0$, then $a_{i_0j_0}a_{kl}=a_{i_0l}a_{kj_0}=0$ would imply that $a_{i_0j_0}=0$, which is a contradiction. Therefore, \eqref{productcondition} holds for all $i,k,l$. Next, we note that $\nu_l^{ik}$ does not depend on $i,k$, because the same multiplicative factor must appear in \eqref{productcondition} when we carry out either of the replacements $i\rightarrow i'$ or $k\rightarrow k'$. We can thus write $\nu_l$ in place of $\nu_l^{ik}$. Finally, we have

\begin{equation}
	\vec{a}=\begin{pmatrix}\vdots\\a_{i1}\\ \vdots\\a_{iN}
	\\ \vdots\\a_{k1}\\ \vdots\\ a_{kN}\\ \vdots
	\end{pmatrix}
	=\begin{pmatrix}\vdots\\a_{ij_0}\begin{pmatrix}\nu_1\\ \vdots\\ \nu_N\end{pmatrix}\\ \vdots\\ a_{kj_0}\begin{pmatrix}\nu_1\\ \vdots\\ \nu_N\end{pmatrix}\\ \vdots\end{pmatrix}=\begin{pmatrix}a_{1j_0}\\ \vdots\\a_{Mj_0}\end{pmatrix}\otimes\begin{pmatrix}\nu_1\\ \vdots\\ \nu_N
	\end{pmatrix},
\end{equation}
showing that $\vec{a}$ is a product vector indeed.

By Lemma \ref{productlemma}, $\ket{\psi}=\chi\ket{\phi_1}+\ket{\phi_2}$ is a product vector iff its components with respect to the Schmidt basis of $\ket{\phi_1}$ satisfy a set of equations that are at most quadratic in $\chi$. Furthermore, since at least two Schmidt coefficients are non-zero, there is at least one non-trivial quadratic equation in this set, namely $(\chi\eta_1+z_{11})(\chi\eta_2+z_{22})=z_{12}z_{21}$. Therefore, there can be at most two distinct solutions for $\chi$, corresponding to at most two product vectors in the range of $\rho_\text{rank-2}$. For $M=N=2$, there is only one quadratic equation to be satisfied. It follows that there is at least one product vector in the range of $\rho_\text{rank-2}$, which had already been established in \cite{sanpera98}. For $M=2, N=3$, there are two linear equations and one quadratic equation in $\chi$ that are non-trivial, and the solution set can be empty. This means that there are qubit subspaces of $\mathbb{C}^2\otimes\mathbb{C}^3$ that do not contain any product state at all. A rank-2 state with such a range space will therefore have only the trivial LSD.

Note that one does not know $\emph{a priori}$ whether $\rho_\text{rank-2}$ is separable or entangled. If it is actually separable, two cases can arise: (i) the spectral decomposition already gives a decomposition of $\rho_\text{rank-2}$ into a convex sum of product states, or (ii) at least one eigenvector is entangled, and carrying out the above procedure gives $\tilde{\varrho}_\text{sep}=\rho_\text{rank-2}$.

\section{Conclusion and outlook}
In this paper, we have expressed the problem of finding the optimal LSD of any bipartite state as a sequence of SDPs. This allows the efficient numerical computation of optimal LSDs. Indeed, we have described how to use the DPS-(PPT) separability criterion to form a sequence of semidefinite relaxations to the optimal LSD problem. The corresponding sequence of solutions to these SDPs provides a sequence of convex decompositions of $\rho$ which has been shown to converge to the true optimal LSD. A complementary sequence of decompositions is obtained if one uses the DPS*-(PPT) criterion instead. In this case, we still have convergence provided $\rho$ has full-rank. We have implemented this scheme numerically and have found that for qudit-qubit states, the degree of separability can be obtained to a good precision with a reasonable amount of computational resources. The introduction of an additional PPT constraint reduces the number of extensions that have to be considered, for a given error margin, but increases the computational cost significantly. The prescription we have provided illustrates the versatility of the DPS-criterion --- it provides a test for separability, a means to compute certain entanglement-related quantities, and as elaborated in this paper, a way to access the optimal LSD of any given state. Additionally, we have cast the optimal LSD problem for qubit-qutrit states as a SDP, and have shown that the remainder in the optimal LSD is maximally entangled in some special cases. Finally, we have provided analytically the optimal LSD of rank-2 states in any dimension. 

We would also like to highlight a curious link between the optimal LSD and the concurrence of a state. For pure states in $\mathbb{C}^M\otimes\mathbb{C}^N$, the concurrence is defined \cite{albeverio01, rungta01} by $C(\ket{\psi})=\sqrt{\frac{M}{M-1}\left(1-\text{tr}\{\rho_A^2\}\right)}$, where $\rho_A=\text{tr}_B\{\ket{\psi}\bra{\psi}\}$. This is then extended to mixed states via the convex roof construction \cite{uhlmann00}, with $C(\rho)$ equal to the minimum average pure state concurrence, taken over all ensemble decompositions of $\rho$. Since the concurrence is a convex function which vanishes on separable states, one has the inequality $C(\rho)\leq(1-\mathcal{S})C(\varsigma_\text{ent})$. Remarkably, equality holds for many classes of two-qubit states, for instance, the rank-2 states \cite{englert02} and the full-rank states with full-rank $\varrho_\text{sep}$ \cite{wellens01}. In other words, the optimal LSD is often also an optimal decomposition in the concurrence sense, even though the defining properties of the two decompositions are quite different. It is not known whether this is a coincidence stemming from properties unique to two-qubit states \cite{vollbrecht00}, or part of a more general relationship applicable to higher dimensions. In any case, $(1-\mathcal{S})C(\varsigma_\text{ent})$ provides an upper bound for the concurrence of $\rho$, which is, in some sense, the best that one can do with a decomposition of $\rho$ into two parts. In fact, the bound holds as long as $C$ is a normalized quantity defined through a convex-roof construction which vanishes on product (and thus separable) states, not necessarily the concurrence. Computing convex roof quantities, unfortunately, involves a very difficult optimization procedure, so even $C(\varsigma_\text{ent})$ is usually not accessible. Nevertheless, the quantity $1-\mathcal{S}$ alone is an entanglement monotone in its own right \cite{karnas01}, and may even serve as a fairly good upper bound for the concurrence of $\rho$. Heuristically, the process of finding the optimal LSD ``concentrates" all the entanglement properties of $\rho$ into $\varsigma_\text{ent}$, whose concurrence is then expected to be close to maximal. This suggests that $(1-\mathcal{S})$ by itself can serve as a fairly good upper bound for the concurrence of $\rho$. Furthermore, in those instances where $(1-\mathcal{S})C(\varsigma_\text{ent})=C(\rho)$ holds, we even have a decomposition realizing the concurrence of $\rho$.

\begin{acknowledgments}
The author wishes to thank Berthold-Georg Englert and Philippe Raynal for valuable discussions, and for reading and correcting this manuscript. Centre for Quantum Technologies is a Research Centre of Excellence funded by Ministry of Education and National Research Foundation of Singapore.
\end{acknowledgments}

\appendix

\section{Finding the $k$-optimal decomposition using SDP}\label{explicitsdp}
The $k$-th optimization problem (without the PPT constraint) takes place over the real vector space of hermitian operators on $\mathcal{H}_A\otimes\mathcal{H}_B^{\otimes k}$. We will denote an operator (not necessarily normalized) on this space by $\bar{\rho}_k$ or $\bar{\sigma}_k$. A feasible $\tilde{\rho}_k=\text{tr}_{B^{k-1}}\{\bar{\rho}_k\}$ in a valid $k$-decomposition $\rho=\tilde{\rho}_k+\tilde{\sigma}_k$ must satisfy the following conditions:
\begin{itemize}
	\item[(i)]{$\bar{\rho}_k\geq0$},
	\item[(ii)]{$\rho-\text{tr}_{B^{k-1}}\{\bar{\rho}_k\}\geq 0$}.
\end{itemize}
In an actual numerical implementation, we may parameterize $\bar{\rho}_k$ by its components with respect to some orthogonal basis. Recall that we only need to consider the symmetric subspace of $\mathcal{H}_B^{\otimes k}$ which has dimension $d=\begin{pmatrix} d_B+k-1\\k \end{pmatrix}$. An obvious choice would be to use the basis $\{\sigma_i\otimes\tau_j\}_{i=0,\ldots,d_A^2-1,j=0,\ldots,d^2-1}$, where $\sigma_0=\mathbb{I}_{d_A}/d_A,\tau_0=\mathbb{I}_d/d$, while $\{\sigma_i\}_{i=1}^{d_A^2-1}$ and $\{\tau_j\}_{j=1}^{d^2-1}$ are the traceless generators of SU($d_A$) and SU($d$) respectively. In this parameterization, $\bar{\rho}_k=\sum_{i,j}{x_{ij}\sigma_i\otimes\tau_j}$. We will also need to compute the reduced matrices $\underline{\tau}_j\equiv\text{tr}_{B^{k-1}}\{\tau_j\}$, which enters in the expression $\tilde{\rho}_k=\sum_{i,j}{x_{ij}\sigma_i\otimes\underline{\tau}_j}$. Note that among the reduced matrices $\sigma_i\otimes\underline{\tau}_j$, only $\sigma_0\otimes\underline{\tau}_0$ has non-vanishing trace (equal to 1). Therefore, in standard form, the corresponding SDP is:
\begin{equation}
	\begin{array}{ll}
	\text{minimize} & \vec{c}^{\,\text{T}}\vec{x}\\
	\text{subject to} & F(\vec{x})\equiv\mathbf{0}_{d_Ad}\oplus\rho\\
										&\qquad\quad+\sum_{i,j}x_{ij}(\sigma_i\otimes\tau_j)\oplus(-\sigma_i\otimes\underline{\tau}_j)\geq 0,
	\end{array}
\end{equation}
where $c_0=-1$. If the PPT constraint is desired, one additionally computes the matrices $\sigma_i\otimes\tau_j^{\text{T}_{B^{\lfloor k/2 \rfloor}}}$ and proceeds in a similar fashion. For computing the $k^*$-optimal decompositions, we replace the reduced matrices $\sigma_i\otimes\underline{\tau}_j$ with suitably perturbed ones, in accordance with the definitions of $S_{(p)}^{*k}$ in \eqref{tildeS}.

\section{Optimal LSD from another angle}
\begin{figure}[ht]
	\includegraphics[width=\linewidth,clip=]{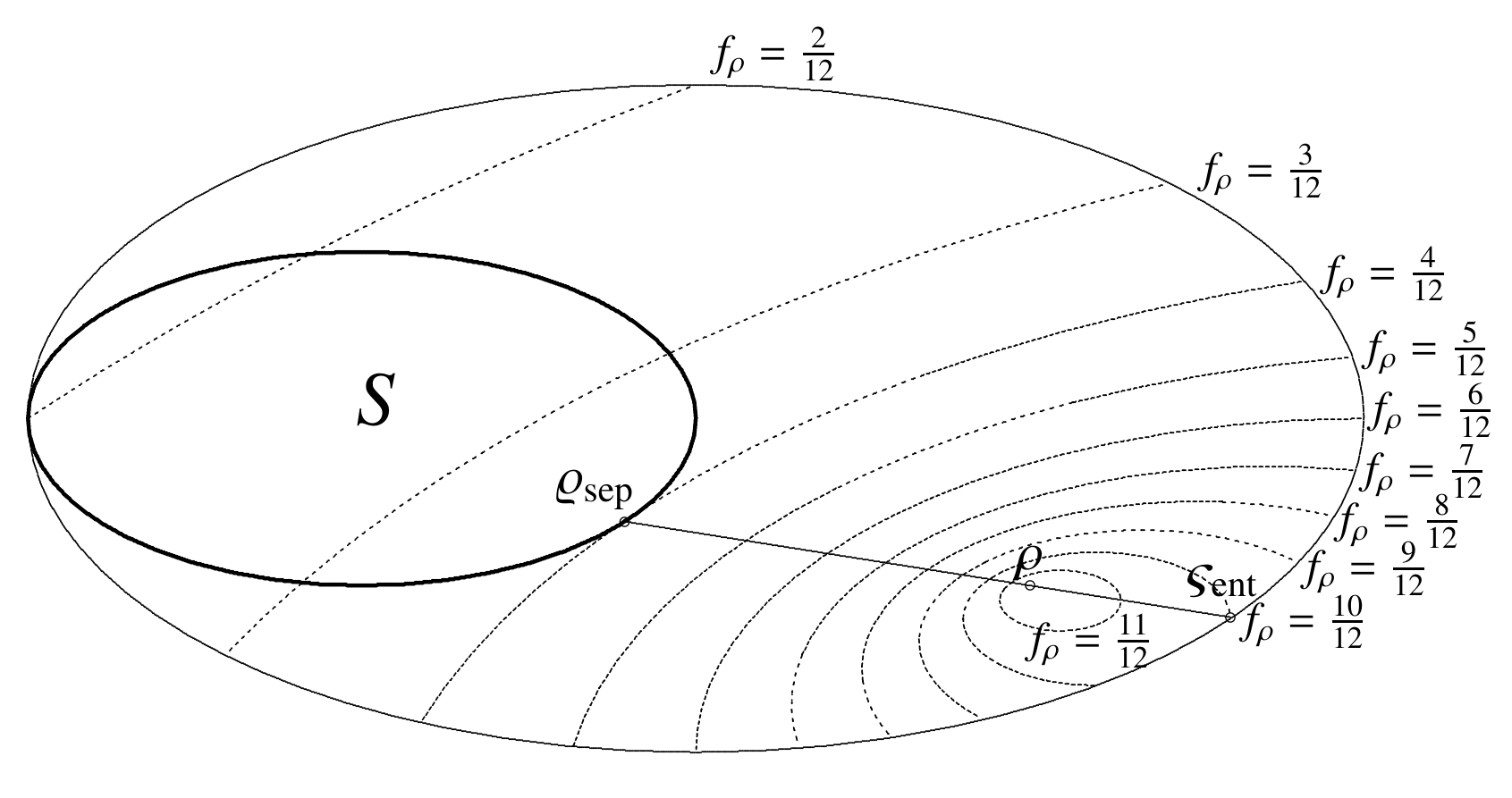}
\caption{Level curves of $f_\rho$ for a full-rank state with $\mathcal{S}=1/3$. The optimal decomposition is illustrated by the straight line. The supporting hyperplane of $S$ at $\varrho_\text{sep}$ also supports the $f_\rho=1/3$ superlevel set.}
\label{levelcurves}

\end{figure}

There is another way to look at the optimal LSD problem, which may be instructive. Let us define a real-valued function $f_\rho$ on the set of normalized states,
\begin{equation}
	f_\rho(\gamma)=\sup\{\lambda:\rho-\lambda\gamma\geq 0\},
\end{equation}
which takes on values in the closed interval $[0,1]$. In particular, one has $f_\rho(\varrho_\text{sep})=\mathcal{S}$. The degree of separability of $\rho$ is then the maximal value of $f_\rho|_S$ restricted to the set of separable states $S$. The argument at this maximum, $\text{arg max}\:f_\rho|_S(\gamma)$, is then $\varrho_\text{sep}$. It is possible to show that $f_\rho$ is a quasiconcave function, that is, whenever, $\gamma_1$ and $\gamma_2$ are two normalized states, and $\theta\in [0,1]$, then
\begin{equation}
	f_\rho\left(\theta\gamma_1+(1-\theta)\gamma_2\right)\geq \min\left(f_\rho(\gamma_1),f_\rho(\gamma_2)\right).\label{qconcave}
\end{equation}
Let us assume that $f_\rho(\gamma_1)$ is smaller. Then
\begin{align}
\rho-f_\rho(\gamma_1)\left[\theta\gamma_1+(1-\theta)\gamma_2\right]&=\theta\left[\rho-f_\rho(\gamma_1)\gamma_1\right]\nonumber\\
	&\;\;\;+(1-\theta)\left[\rho-f_\rho(\gamma_2)\gamma_2\right]\nonumber\\
	&\;\;\;+(1-\theta)\left[f_\rho(\gamma_2)-f_\rho(\gamma_1)\right]\gamma_2\nonumber\\
	&\geq 0,
\end{align}
from which it follows that $f_\rho\left(\theta\gamma_1+(1-\theta)\gamma_2\right)\geq f_\rho(\gamma_1)$, i.e., \eqref{qconcave} holds. Now, an equivalent definition of quasiconcavity is that every $\alpha$-superlevel set $\{\gamma\: :\: f_\rho(\gamma)\geq \alpha\}$ is convex. This gives a nice geometrical picture of our problem using the level curves of $f_\rho$ (Fig.~\ref{levelcurves}).

\end{document}